\documentclass{article}
\newcommand{\nazevclanku}{Weyl metrics and the generating conjecture}
\newcommand{\autori}{L. Richterek and J. Horsk\'y}
\usepackage{cjp}
\usepackage{graphicx,richpom}
\begin{document}
\title{Weyl metrics and the generating conjecture}
\author{%
L. Richterek\footnote{E-mail: {\tt richter@aix.upol.cz}}\\
    {\sl Department of Theoretical Physics, Palack\'y University,}\\
    {\sl 17. listopadu 50, Olomouc, 772 00}\\[\medskipamount]
J. Horsk\'y\footnote{E-mail: {\tt horsky@physics.muni.cz}}\\   
    {\sl Institute of Theoretical Physics and Astrophysics, Masaryk University,}\\ 
    {\sl Kotl\'a\v{r}sk\'a 2, Brno, 611 37, Czech Republic}
}\maketitle

\begin{abstract}
By means of Ernst complex potential formalism it is shown, that previously
studied static axisymmetric Einstein-Maxwell fields obtained though the
application of the Horsk\'y-Mitskievitch generating conjecture represent a
combination of Kinnersley's transformations [W.~Kinnersley:
J. Math. Phys. \textbf{14} (1973) 651]. New theoretical background for the
conjecture is suggested and commented.
\end{abstract}


\section{Introduction}
\label{sec:rhintro}

The complexity of Einstein and coupled Einstein-Maxwell (EM) equations in
general relativity led many authors to invent a lot of generating
techniques that enable us to obtain new families of EM solutions from those
already known (see e.g.~\cite{ksmh}, \S 30.1--\S 30.6 for a compact
overview) instead of solving the field equations. Most of these techniques
require or employ spacetime symmetries in some way which allows to simplify
the problem to a certain extent. One of such methods demanding the presence
of at least one Killing vector for the seed metric was proposed
in~\cite{Horsky:1989}. Though the application of the Horsk\'y-Mitskievitch
(HM) conjecture have resulted in finding several classes of new EM
fields~\cite{Horsky:1990,Cataldo:1994,Cataldo:1995,Fikar:1999,Klepac:2000,Richterek:2000,Richterek:2002},
the conjecture itself has not been proved so far. Our objective is to
contribute to a deeper understanding of the HM conjecture and to explain
its possible connections with other more thoroughly explored generating
methods.

The paper is organized as follows. In Section~\ref{sec:rhsymEM} we
summarize basic facts about complex potential formalism developed by
Ernst~\cite{Ernst:1968a,Ernst:1968b}, Section~\ref{sec:rhexamp} is devoted
to particular examples of Harrison transformations applied to the EM fields
generated by means of HM conjecture from some classes of the Weyl vacuum
metrics, in Section~\ref{sec:rhconj} we compare both approaches and discuss
the common features of these generating techniques. 

Throughout the text geometrized units in which $c=1$, $G=1$ are used. The metric
signature $-+++$ and the indexing conventions follow~\cite{Carroll:2003}.

\section{Symmetries of stationary Einstein-Maxwell fields}
\label{sec:rhsymEM}
Though the EM equations describing the coupling of the electromagnetic field
with gravity are very complicated, it has been demonstrated they are
endowed with hidden symmetries. The groups of corresponding transformations
for stationary EM spacetimes were systematically described by Kinnersley
and his coworkers
(see~\cite{Kinnersley:1973,Kinnersley:1977pg,Kinnersley:1980kj} and
references cited therein).

Following Kinnersley, in this paper we also make use of complex potential formalism
designed by Ernst~\cite{Ernst:1968a,Ernst:1968b} for an effective description
of stationary axially symmetric EM fields, namely the Kerr and Kerr-Newman
solutions. Afterwards it was generalized for stationary spacetimes by
Ernst~\cite{Ernst:1971} and independently by Israel and Wilson~\cite{Israel:1972}.
Let us recall the main ideas and necessary formulae.
A general stationary line-element may be written in the form
\begin{equation}
{\rm d}s^2=-f\left({\rm d}t-\omega_j{\rm d}x^j\right)^2 +
\frac{1}{f}\,h_{jk}\,{\rm d}x^j{\rm d}x^k,\quad j,k=1,2,3
\label{eq:rhstacmet}
\end{equation}
where $\{x^j\}$ represent some spacelike coordinates and metric functions $f,\
 w_j,\ h_{jk}$ do \emph{not} depend on $t$. Denoting the covariant
 derivative with respect to the 3-dimensional metric $h_{jk}$ as $\nabla$,
 one can define a twist vector
\[
\bgchi=f^2\nabla\times\bgomega+\,{\rm i}\left(
\Phi^{\ast}\nabla\Phi-\Phi\nabla\Phi^{\ast}\right),\qquad
{\rm i}=\sqrt{-1}
\]
satisfying the equation
\[
\nabla\times\bgchi=0,
\]
which implies the existence of a scalar
``twist potential'' $\psi$, such that
\begin{equation}
\bgchi = \nabla\psi.
\label{eq:rhtwpot}
\end{equation}
If we introduce a complex scalar potential $\Phi$ describing the electromagnetic
field (see below) and another complex scalar ${\cal E}$ by the relation
\begin{equation}
{\cal E}=f-\left|\Phi\right|^2+{\rm i}\psi,
\label{eq:rhernstpot}
\end{equation}
then the system of coupled EM equations may be replaced by two
complex 3-dimensional Ernst equations for Ernst potentials ${\cal E}$,
$\Phi$
\begin{equation}
\eqalign{%
f\Delta{\cal E}=&\left(\nabla{\cal E}+2\Phi^{\ast}\nabla\Phi\right)\cdot\nabla{\cal E},\cr
f\Delta\Phi=&\left(\nabla{\cal E}+2\Phi^{\ast}\nabla\Phi\right)\cdot\nabla\Phi.}
\label{eq:rhernsteq}
\end{equation}
It is possible to classify various types of stationary EM fields according to
the values taken by the complex Ernst potentials ${\cal E}$ and $\Phi$ as
it is summarized in Table~\ref{tab:rhtypyspt}~\cite{ksmh}.
\begin{table}[htbp]
\centering
\caption{Ernst potentials for various types of stationary EM fields}
\label{tab:rhtypyspt}
\begin{tabular}{|c|c|c|}
\hline
\emph{spacetime}&$\cal{E}$&$\Phi$\\
\hline
stationary EM fields& complex & complex\\
static electrovac fields& real & real\\
static magnetovac fields& real & imaginary\\
stationary vacuum field& complex & $0$\\
static vacuum field& real & $0$\\
conform-stationary EM fields&$0$ & complex\\
\hline
\end{tabular}
\end{table}

For the sake of simplicity and without loss of generality we may further
consider a timelike electromagnetic vector potential
$\bvec{A}=A_t\bvec{dt}$ corresponding to an electric field. In this case the
Ernst potential $\Phi=A_t$ reads straightforwardly, while in magnetic case $\Phi$
gets imaginary values and has to be found as a solution of partial differential
equations (see e.g.~\cite{Ernst:1968b,Richterek:2002} for
examples). Moreover, we can turn electric field into magnetic and vice versa
via duality rotation in the complex plane of the potential $\Phi$.

Kinnersley~\cite{Kinnersley:1973,Kinnersley:1977pg} proved, that the EM
equations in the presence of one non-null Killing vector posses
covariance under an 8-parameter group of transformation isomorphic to
$SU(2,1)$. Those eight parameters can be combined into three complex parameters
$a,b,c$ and two real parameters $\alpha,\beta$, so that 
from a given solution it is possible to generate a five-parameter family of
solutions; the change of Ernst potentials under the symmetry
transformations is summarized in Table~\ref{tab:rhkintr}. Lets us remind
that the spacelike metric coefficients $h_{ik}$ in~(\ref{eq:rhstacmet})
remain unchanged.
\begin{table}[htbp]
\centering
\caption{Kinnersley transformations: $a,b,c\in C$, $\alpha,\,\beta\in R$}
\label{tab:rhkintr}
\begin{tabular}{|p{0.42\linewidth}|l|}
\hline
I: Electromagnetic gauge&\parbox[c][14ex]{0.45\linewidth}{%
\begin{equation}
\eqalign{%
    {\cal E}&\longrightarrow {\cal E}-2a^{\ast}\Phi-aa^{\ast},\cr
    \Phi&\longrightarrow \Phi+a,\cr
    f&\longrightarrow f,\quad \bgchi\longrightarrow\bgchi
}
\label{eq:rhelmgauge}
\end{equation}
}
\\
\hline
II: Gravitational gauge&\parbox[c][14ex]{0.45\linewidth}{%
\begin{equation}
\eqalign{%
    {\cal E}&\longrightarrow {\cal E}+{\rm i}\alpha,\cr 
    \Phi&\longrightarrow \Phi,\cr
    f&\longrightarrow f,\quad \bgchi\longrightarrow\bgchi
}\label{eq:rhgrgauge}
\end{equation}       
}\\
\hline
III: $\cases{|b|=1&\cr
             \mbox{duality rotation}&\cr
             &\cr
             |b|\neq1&\cr
             \mbox{scaling or conformal transf.} 
     }$
     &\parbox[c][16ex]{0.45\linewidth}{%
\begin{equation}
\eqalign{%
    {\cal E}&\longrightarrow \left(bb^{\ast}\right)^{-1}{\cal E},\cr
    \Phi&\longrightarrow \left(b^{\ast}b^{-2}\right)\Phi,\cr
    {\rm d}s^{2}&\longrightarrow \left(bb^{\ast}\right)^{-1}{\rm d}s^{2}
}\label{eq:rhdualrot}
\end{equation}   
}\\
\hline
IV: Ehlers transformation&\parbox[c][14ex]{0.45\linewidth}{%
\begin{equation}
\eqalign{%
{\cal E}&\longrightarrow\frac{{\cal E}}{1+{\rm i}\beta{\cal E}},\cr
\Phi&\longrightarrow\frac{\Phi}{1+{\rm i}\beta{\cal E}}
}\label{eq:rhehl}
\end{equation}    
}\\
\hline
V: Harrison transformation&\parbox[c][14ex]{0.45\linewidth}{%
\begin{equation}
\eqalign{%
  {\cal E}&\longrightarrow\frac{{\cal E}}{1-2c^{\ast}\Phi-cc^{\ast}{\cal E}},\cr
  \Phi&\longrightarrow\frac{\Phi+c{\cal E}}{1-2c^{\ast}\Phi-cc^{\ast}{\cal E}}
}
\label{eq:rhhar}
\end{equation}
}\\
\hline
\end{tabular}
\end{table}

The gauge transformations~(\ref{eq:rhelmgauge}) and~(\ref{eq:rhgrgauge})
are of course not interesting from a physical point of view, as they do not
lead to a new metric. The duality rotation~(\ref{eq:rhdualrot}) for a
complex unit parameter $|b|=1$ changes the type of electromagnetic field
not altering the metric line element. The Ehlers
transformation~(\ref{eq:rhehl}) reverts static fields into stationary ones
and its more general Kinnersley's form in Table~\ref{tab:rhkintr} admits
the presence of the electromagnetic field unlike its original Ehlers
formulation~\cite{Kinnersley:1977pg}. Finally, the Harrison
transformation~(\ref{eq:rhhar}) may add an electromagnetic field with
$\Phi\neq 0$ to a vacuum seed metric for which $\Phi=0$. It is namely this
``charging'' transformation we would like to concentrate on in comparison
with the HM generating conjecture.

We can see, that the existence of a single non-null Killing vector field
endows the EM field with a remarkable amount of symmetry and internal
structure described above.  Naturally, the situation becomes considerably
simpler in the presence of two commuting Killing vectors as in the
frequently studied case of stationary axisymmetric fields. Most examples
described in Section~\ref{sec:rhexamp} belong to this class. The
effectiveness of the complex potential approach was systematically
demonstrated in a series of papers by Hauser and Ernst and completed by
their proof of a generalized Geroch conjecture~\cite{Hauser:2001}. Thus,
all vacuum spinning mass solutions could be generated from any one such
solution (even Minkowski space) by means of an infinite sequence of
transformations associated with the internal symmetries and with the choice
of the basic Killing vector fields.

\section{Examples of Kinnersley's transformations}
\label{sec:rhexamp}
In this section we demonstrate, that some EM fields we derived via HM
generating conjecture represent either Harrison transformation or a
combination of Kinnersley's
transformations~(\ref{eq:rhelmgauge})--(\ref{eq:rhhar}) of the
corresponding seed metrics.  We are going to concentrate on electro- and
magnetovacuum solutions obtained from seed axially symmetric vacuum
gravitational fields~\cite{Richterek:2000,Richterek:2002}. For the static
EM fields the twist potential $\psi$ in~(\ref{eq:rhtwpot}) equals zero and
the potential ${\cal E}$ takes real values only (see
Table~\ref{tab:rhtypyspt}). According to~(\ref{eq:rhernstpot}) the Ernst
potentials read as
\begin{equation}
  \label{eq:rhetr}
  {\cal E}_{\rm seed}=f_{\rm seed},\qquad
  \Phi_{\rm seed}=0,
\end{equation}
for the seed metrics and 
\begin{equation}
  \label{eq:rhetrfm}
  {\cal E}_{\rm charged}=f_{\rm charged}-\left|\Phi_{\rm charged}\right|^2,\qquad
  \Phi_{\rm charged}\neq0,
\end{equation}
for the charged solutions provided they remain static. Moreover, for studied
electrovacuum solutions with a timelike vector potential we can just put
\begin{equation}
  \label{eq:rhphitr}
  \Phi=A_t.
\end{equation}
Despite of relative simplicity, the Weyl solutions class includes many
astrophysically interesting solutions that might be relevant e.g. for the
description of gravitating discs around black
holes~\cite{bicak:2000,Semerak:2002}.

All the solutions found by the authors employing HM generating conjecture
revealed that for a particular class of charged solutions the metric
coefficients were modified in the same way, no matter whether we start with
the Levi-Civita~\cite{Richterek:2000} or the Darmois-Vorhees-Zipoy
(also $\gamma$)~\cite{Richterek:2002} metrics. Let us show, that this
modification represents the Harrison transformation~(\ref{eq:rhhar}) in
fact.

Let us start with the $\gamma$-metric with an electric field, the line
element and fourpotential of which in the Weyl-Lewis-Papapetrou cylindrical
coordinates read as
\begin{equation}
\eqalign{%
{\rm d}s^{2}&= -\,\frac{f_{1}(r,z)}{f(r,z)^2}\,{\rm d}t^{2} + 
\frac{f(r,z)^2}{f_{1}(r,z)}\,
\left[f_{2}(r,z)\left({\rm d}r^{2}+{\rm d}z^{2}\right) + 
r^{2}\,{\rm d}\varphi^2\right],\cr
\bvec{A}&=q\,\frac{f_{1}(r,z)}{f(r,z)}\,\bvec{dt},
}
\label{eq:rhgamel}
\end{equation}
where
\begin{eqnarray*}
f(r,z)&=&1-q^2f_{1}(r,z),\cr
f_1(r,z)&=&\left(\frac{R_{1}+R_{2}-2m}{R_{1}+R_{2}+2m}\right)^{\gamma},\quad
f_2(r,z)=\left[\frac{(R_{1}+R_{2}-2m)(R_{1}+R_{2}+2m)}
        {4R_{1}R_{2}}\right]^{\gamma^{2}},\cr
R_{1}&=&\sqrt{r^{2}+\left(z-m\right)^{2}},\qquad
R_{2}=\sqrt{r^{2}+\left(z+m\right)^{2}}.
\end{eqnarray*}
Setting $q=0$ we obtain the seed $\gamma$-metric with ${\cal E}_{\rm seed}=f_{1}(r,z)$.
Extracting the Ernst potentials according to~(\ref{eq:rhetr}) and~(\ref{eq:rhphitr}) 
one come to
\[
{\cal E}=\frac{f_{1}(r,z)}{f(r,z)}=\frac{f_{1}(r,z)}{1-q^2f_{1}(r,z)},\qquad
\Phi=q{\cal E}=\frac{qf_{1}(r,z)}{1-q^2f_{1}(r,z)},
\]
which coincides with~(\ref{eq:rhhar}) for a real parameter $c=q$.

In case of solutions with magnetic field we cannot just use the
relation~(\ref{eq:rhphitr}) for $\Phi$, moreover, from the
Table~\ref{tab:rhtypyspt} we know, that $\Phi$ takes imaginary values. 
Rewriting the magnetovacuum $\gamma$-metric line element
\begin{equation}
\eqalign{%
{\rm d}s^{2}=&-\,f(r,z)^2f_{1}(r,z)\,{\rm d}t^{2} +\cr
&+\frac{1}{f_{1}(r,z)}
\left[f(r,z)^2f_{2}(r,z)\left({\rm d}r^{2}+{\rm d}z^{2}\right) + 
\frac{r^{2}}{f(r,z)^2}\,{\rm d}\varphi^2\right]
}
\label{eq:rhgammagorig}
\end{equation}
into form
\[
{\rm d}s^{2}=\frac{r^{2}}{f(r,z)^2f_{1}(r,z)}\,{\rm d}\varphi^2+
f(r,z)^2f_{1}(r,z)\left[\frac{f_{2}(r,z)}{f_{1}(r,z)^2}\left({\rm d}r^{2}+
{\rm d}z^{2}\right)-\,{\rm d}t^{2}\right]
\]
with
\[
\bvec{A}=q\,\frac{r^2}{f(r,z)f_{1}(r,z)}\,\bvec{d}\bgvarphi,\qquad
f(r,z)=1+q^2\frac{r^2}{f_{1}(r,z)},
\]
we realize, that a formal transformation interchanging $t$ and
$\phi$ coordinates makes the situation mathematically equivalent to the
electrovacuum solution~(\ref{eq:rhgamel}) with ${\cal E}_{\rm
seed}=-r^2/f_{1}(r,z)$. Indeed, the both Ernst potentials of the ``charged''
solution
\[
{\cal E}=\frac{-\,\displaystyle\frac{r^2}{f_{1}(r,z)}}
{1-q^2\left[-\,\displaystyle\frac{r^2}{f_{1}(r,z)}\right]},\quad
\Phi=\frac{-\,q\,\displaystyle\frac{r^2}{f_{1}(r,z)}}
{1-q^2\left[-\,\displaystyle\frac{r^2}{f_{1}(r,z)}\right]}
\]
again fulfil the Harrison transformation equations~(\ref{eq:rhhar}).

The same could be gradually accomplished for {\em all} the solutions
generated from the Levi-Civita seed metric in~\cite{Richterek:2000} -- they
represent its Harrison transformation~(\ref{eq:rhhar}). Naturally, one
would like to check the spacetimes studied in~\cite{Horsky:1989} to support
the HM conjecture. The first example -- the charged Schwarzschild and
Kerr, i.e. the Reissner-Nordstr\"om and Kerr-Newman solutions respectively
-- has been proved to be a combination of Kinnersley's transformations in
fact by Ernst~\cite{Ernst:1968b}. The second example in~\cite{Horsky:1989}
-- the charged Taub solution
\begin{equation}
{\rm d}s^{2}=-\frac{a+bx}{x^2}\,{\rm d}t^{2} +\frac{x^2}{a+bx}\,{\rm d}x^{2}+
x^2\left[{\rm d}y^{2} + {\rm d}z^{2}\right],\qquad
\bvec{A}\sim\frac{b}{x}\,\bvec{dt}
\label{eq:rheltaub}
\end{equation}
reduces to the Taub solution for $a=0$. Thus, according to~(\ref{eq:rhetr})
${\cal E}_{\rm seed}=b/x$. The Harrison transformation~(\ref{eq:rhhar})
then gives
\[
\Phi=\frac{cb/x}{1-c^2b/x},\quad
{\cal E}=\frac{b/x}{1-c^2b/x},\quad 
f=\mbox{Re}{\cal E}+\left|\Phi\right|^2=\frac{bx}{\left(x-bc^2\right)^2}.
\]
Introducing a new spacelike coordinate $\xi$ by the relations
\[
x=\xi+bc^2,\qquad a=b^2c^2
\]
we get the metric~(\ref{eq:rheltaub}).

We have explicitly demonstrated that the HM conjecture in case of some static
fields provides results equivalent to Harrison transformation. The common
issues of both methods are discussed in the following section.

\section{The generating conjecture in a new context}
\label{sec:rhconj}
Having been proposed in~\cite{Horsky:1989}, the HM conjecture was
reformulated to meet more suitably practical generation of new EM
fields. Originally, the conjecture says, that having a seed vacuum
gravitational field with at least one Killing vector, then it makes sense
to search for an EM spacetime for which the fourpotential of the
electromagnetic field is proportional (up to a constant factor) to the
Killing covector of the seed vacuum metric. When the parameter connected
with the electromagnetic field of the self-consistent problem is set equal
to zero, one comes back to the seed solution. After several successful
application of the conjecture it was generalized by Cataldo et
al.~\cite{Cataldo:1994} in the sense, that the electromagnetic
fourpotential need not be just a constant multiple of the seed metric
Killing covector, but that it is possible to multiply by a suitable
function. Finally, the conjecture was also used in a few cases when the
seed spacetimes were non-vacuum solutions of the Einstein
equations~\cite{Horsky:1990,Cataldo:1995}. Thus the key idea is that the
electromagnetic field tensor $\bvec{F}=\bvec{dA}$ is in some sense connected
with so called Papapetrou fields~\cite{Fayos:2002} -- exterior
derivatives of corresponding Killing fields.

The conjecture does not specify, in what way the metric tensor of the EM
field is modified in comparison with the seed vacuum spacetime, thus it does not
provide an exact algorithm, how to generate charged solutions from the seed
ones. Unfortunately, in many cases it is extremely difficult (if not even
impossible) to solve EM equations without any additional condition, even
if we set the electromagnetic fourpotential in accordance with the
conjecture.

There is only one condition required by the conjecture: the existence of a
Killing vector field. The conjecture does not impose any restriction,
whether the Killing vector should be timelike, spacelike, null or whether
some Killing vectors should be excluded.  The possible connection of the
electromagnetic field with spacetime symmetries described by the Killing
vectors is also not closely specified.

On the other hand, we have demonstrated in Section~\ref{sec:rhexamp} that
some classes of the EM fields found by means of the conjecture are in fact
examples of Kinnersley's transformations. Similarly, the usage of these
transformations demands an existence of a non-null Killing
vector. Moreover, the charging Harrison transformation prescribes
exactly, how to modify Ernst potentials of the seed metric and thus
provides generating algorithm of consequent calculations. And finally, the
Harrison transformation is also connected with the concept of symmetry: a
non-null Killing vector is needed for the 8-parameter group of
transformations described in Section~\ref{sec:rhsymEM}.

Conversely, for all classes of vacuum space time admitting a non-null
Killing vector the Harrison transformation ensures an existence of a
correspondent charged EM field and supplies a procedure, how to construct
it. From this point of view for all solutions generated from the seed Weyl
metrics in~\cite{Richterek:2000,Richterek:2002} the HM conjecture in its
generalized formulation~\cite{Cataldo:1994} necessarily had to work and
provide new EM fields. Of course, the Harrison transformation is really
simple for static metrics in Section~\ref{sec:rhexamp}, where we do not
need to take into account the zero twist potential $\psi$
in~(\ref{eq:rhernstpot}). Anyway, the Kinnersley
transformations~(\ref{eq:rhelmgauge})--(\ref{eq:rhhar}) explain the
validity of the HM conjecture for a wide class of seed metrics.

\section{Conclusions}
\label{sec:rhconcl}
The connection between the HM conjecture and inner symmetries of EM fields
described in special case by Kinnersley's transformations as suggested in
the preceding section might give a more solid theoretical background to
the conjecture and could lead to its more precise formulation or even
explanation. Let us remind, the HM conjecture in connection with complex
potentials has been considered by Stephani~\cite{stephani:2000} who
explored the original formulation with a fourpotential being a constant
multiple of a corresponding Killing vector. He has proved the HM
conjecture for some class of EM fields admitting a diverging, geodesic and
shearfree null congruence and with a non-radiative Maxwell field.

The possible connection of the HM conjecture with inner symmetries of the
 EM equations proposed above would connect the conjecture with a set of
 generating methods elaborated by Ernst and other authors (see
 e.g.~\cite{Ernst:1994jf,Hauser:2001} and references cited therein) for
 axisymmetric fields. These methods are based on the solution of the
 homogeneous Hilbert problem for the axes-accessible Einstein equations
 (solutions with singularities along the whole axis such as Levi-Civita's
 one are excluded) and it was proved~\cite{Hauser:2001}, that these vacuum
 fields are deducible through the action of a huge group with infinitesimal
 generators. It turns out that the axis values of $\cal{E}$ contain enough
 information to construct Ernst potentials at off-axis points, the axis
 mass distribution, angular momentum, electric and magnetic charge
 distributions.  Such axis relation provides a way to identify a
 corresponding Kinnersley-Chitre transformation to generate a spacetime
 with prescribed ${\cal E}$ potential from Minkowski space via this Geroch
 group. The application of this group covers physically interesting
 problems such as derivation of the Kerr metric, spinning-mass solutions of
 arbitrary complexity, the cylindrical gravitational wave and the colliding
 plane gravitational waves solution. Moreover, the proposed connection of
 the HM conjecture with Kinnersley's transformations would suit Stephani's
 demand~\cite{stephani:2000} on its invariant formulation.

Naturally, there still remain open problems. It is necessary to check other
EM fields generated through the conjecture, especially those with
non-static or non-vacuum seed metrics. The Kinnersley transformations does
not support the HM conjecture employing null Killing vectors.

We believe that the HM conjecture reflects some hidden principles. The
Kinnersley's transformation may represent a right clue to its better understanding.

{\bigskip\small%
The authors are obliged to Prof. Jan~Novotn\'y (Dept. of General Physics,
Fac. of. Science, Masaryk University, Brno) for helpful discussions and his
critical comments. They also gratefully appreciate the information and
texts provided by Prof.~F.J.~Ernst on FJE Enterprises' website
\texttt{http://pages.slic.com/gravity/}.}

\bibliography{richor04}
\bibliographystyle{cjp}
\end{document}